\documentstyle[twoside,fleqn,espcrc2]{article}

\newcommand{\AmS}{{\protect\the\textfont2
  A\kern-.1667em\lower.5ex\hbox{M}\kern-.125emS}}

\def\ra{\!\!\rightarrow\!\!}
\def\ol{\prime}
\def\half{{\scriptstyle{1\over 2}}}

\def\tw{{\scriptstyle{1\over 12}}}
\def\quart{{\scriptstyle{1\over 4}}}
\def\zahlen{{\rm Z \!\! Z}}
\def\sn{{\scriptstyle{1-}}}
\def\Tr{{\rm Tr}}
\def\cO{{\cal O}}
\def\cF{{\cal F}}
\def\cD{{\cal D}}
\def\cZ{{\cal Z}}

\def\be{\begin{equation}}
\def\ee{\end{equation}}
\def\bea{\begin{eqnarray}}
\def\eea{\end{eqnarray}}
\def\Journal#1#2#3#4{{#1} {#2}, #3 (#4)}
\def\NPB{{Nucl. Phys.} B}
\def\NPBp{{Nucl.Phys.} B}
\def\PLB{{Phys. Lett.} B}

\def\CMP{Comm. Math. Phys.}
\def\PRD{{Phys. Rev.} D}

\newcommand{\basispl}{
   \put(-.5,-.5){\line(1,0){1}}
   \put(.5,-.5){\line(0,1){1}}
   \put(.5,.5){\line(-1,0){1}}
   \put(-.5,.5){\line(0,-1){1}}}

\newcommand{\twooneplaq}{\setlength{\unitlength}{.5cm}
   \raisebox{-.2cm}{
   \begin{picture}(2.2,1.2)(-1.1,-.6)
   \put(-1,-.5){\line(1,0){2}}
   \put(-1,.5){\line(1,0){2}}
   \put(-1,-.5){\line(0,1){1}}
   \put(1,-.5){\line(0,1){1}}
   \multiput(-1,-.5)(1,0){3}{\circle*{.2}}
   \multiput(-1,.5)(1,0){3}{\circle*{.2}}
   \end{picture}}}
\newcommand{\plaqa}{\setlength{\unitlength}{.5cm}\raisebox{-.2cm}{
   \begin{picture}(1.2,1.2)(-.6,-.6)
   \basispl
   \put(-.5,-.5){\circle*{.2}}
   \put(-.5,.5){\circle*{.2}}
   \put(.5,-.5){\circle*{.2}}
   \put(.5,.5){\circle*{.2}}
   \end{picture}}}
\newcommand{\hookplaq}{\setlength{\unitlength}{.5cm}
   \raisebox{-.3268cm}{
   \begin{picture}(1.7071,1.7071)(-.7071,-.7071)
   \put(0,0){\line(0,1){1}}
   \put(0,1){\line(1,0){1}}
   \put(1,1){\line(0,-1){1}}
   \put(-.7071,-.7071){\line(1,0){1}}
   \put(0,0){\line(-1,-1){.7071}}
   \put(1,0){\line(-1,-1){.7071}}
   \multiput(0,0)(1,0){2}{\circle*{.2}}
   \multiput(0,1)(1,0){2}{\circle*{.2}}
   \multiput(-.7071,-.7071)(1,0){2}{\circle*{.2}}
   \multiput(0,0)(.25,0){4}{\circle*{.03}}
   \end{picture}}}
\newcommand{\cornplaq}{\setlength{\unitlength}{.5cm}
   \raisebox{-.3268cm}{
   \begin{picture}(1.7071,1.7071)(-.7071,-.7071)
   \put(-.7071,-.7071){\line(0,1){1}}
   \put(0,1){\line(1,0){1}}
   \put(1,1){\line(0,-1){1}}
   \put(-.7071,-.7071){\line(1,0){1}}
   \put(0,1){\line(-1,-1){.7071}}
   \put(1,0){\line(-1,-1){.7071}}
   \put(-.7071,-.7071){\circle*{.1}}
   \put(-.7071,.2929){\circle*{.2}}
   \multiput(0,0)(1,0){2}{\circle*{.2}}
   \multiput(0,1)(1,0){2}{\circle*{.2}}
   \multiput(-.7071,-.7071)(1,0){2}{\circle*{.2}}
   \multiput(0,0)(.25,0){4}{\circle*{.03}}
   \multiput(0,0)(0,.25){4}{\circle*{.03}}
   \multiput(0,0)(-.1768,-.1768){4}{\circle*{.03}}
   \end{picture}}}
\newcommand{\twoplaq}{\setlength{\unitlength}{1cm}\raisebox{-.5cm}{
   \begin{picture}(1.2,1.2)(-.6,-.6)
   \basispl
   \put(-.5,-.5){\circle*{.1}}
   \put(-.5,.5){\circle*{.1}}
   \put(.5,-.5){\circle*{.1}}
   \put(.5,.5){\circle*{.1}}
   \put(0,-.5){\circle*{.1}}
   \put(0,.5){\circle*{.1}}
   \put(.5,0){\circle*{.1}}
   \put(-.5,0){\circle*{.1}}
   \end{picture}}}

\title{
Improved action and Hamiltonian in finite volumes\thanks{Based on the 
talks ``Testing Improvement'' and ``Hamiltonian from Improved Action'' by the 
last two authors at Lattice'96, St. Louis, 4-8 June 1996.}}
\author{Margarita Garc\'{\i}a P\'erez, Jeroen Snippe and Pierre van Baal
\address{Instituut-Lorentz for Theoretical Physics, University of Leiden,\\
 PO Box 9506, NL-2300 RA Leiden, The Netherlands}}
\begin{document}
\begin{abstract}
We introduce a new Symanzik improved action by adding a $2\times 2$ plaquette 
in such a way that the Feynman rules in the covariant gauge simplify. We call
this the square Symanzik action. Some comparisons with the continuum and the 
standard Wilson action are made in intermediate volumes, where mass ratios 
are accurately known and the precise amount of improvement can be determined. 
Ratios of the Lambda parameters will be presented, as well as partial results 
for the one-loop improvement coefficients. We discuss some of the intricacies 
that arise because of violations of unitarity at the scale of the cutoff. In 
particular we show how a field redefinition in the zero-momentum effective 
action allows one to remove scaling violations linear in the lattice spacing. 
\end{abstract}
\maketitle
\section{INTRODUCTION}
We consider here the Symanzik improvement scheme~\cite{sym}, which is designed 
to remove lattice artefacts by adding irrelevant operators to the lattice 
action, whose coefficients are tuned by requiring spectral quantities to be 
improved to the relevant order (on-shell improvement~\cite{luwe,onsh}). 
Perturbative calculations, although difficult, are still manageable. For 
Symanzik improvement to work it seemed that unreasonably small values of the 
bare coupling constant were required. 

Mean field inspired Symanzik improvement~\cite{par,lema} was introduced 
to beat the bad convergence of perturbation expansions in the bare coupling 
constant. In particular the Parisi mean field coupling~\cite{par} defined in 
terms of the plaquette expectation value is seen to improve considerably the 
approach to asymptotic scaling. Despite some attempts~\cite{per} no good 
theoretical understanding for this is available. In addition the prescription
is argued to include tadpole corrections to the coefficients in the Symanzik 
improved action, which can be seen as a mean field renormalization of the link 
variables on the lattice. Only phenomenological arguments have been provided 
to support this. 

One difficulty in testing improvement is how to determine to which extent 
improvement has actually been achieved. For pure gauge theories standard tests 
involve restoration of rotational invariance in the heavy quark 
potential~\cite{lema}.  It becomes more problematic when one has to base the 
judgement on carefully extrapolated Wilson data.

These problems inspired us to consider testing improvement for the pure gauge
glueball spectrum in intermediate volumes, particularly emphasizing the need
to test improvement of scaling. In spectroscopy asymptotic scaling is not such
an important issue since one has to set the scale by fixing one of the masses
anyhow. The main reason for considering intermediate volumes (up to 0.75 fermi 
across) is that this volume range can be accurately described in terms of an 
effective zero-momentum model, nevertheless incorporating important 
non-perturbative features that contribute to energy of electric flux. For SU(2)
results are known both for the continuum limit and for the Wilson lattice 
action, from which precise statements on the scaling violations for the mass 
ratios can be made. 

\section{SQUARE SYMANZIK ACTION}
As usual, one connects a continuum configuration with one on the lattice
by parallel transport of the vector potential along the links 
\be
U_\mu(x)=P\exp(\int_0^a A_\mu(x+s\hat\mu)ds)\ .
\ee
This allows one to extract the irrelevant higher order operators that 
need to be cancelled in a lattice action. 
We introduce the new class of actions by adding a $2\times2$ plaquette to 
the ones considered by L\"uscher and Weisz,
\bea
&&\hskip-8mm S(\{c_i\})\equiv\sum_x\Tr\{c_0\!\!\left\langle\!\sn\plaqa\,\right
\rangle\!+\!2c_1\!\!\left\langle\!\sn\twooneplaq\,\right\rangle\!+\nonumber\\&&
\hskip-8mm\frac{4}{3}c_2\!\left\langle\!\sn\cornplaq \,\right\rangle\!\!+\!4c_3
\!\left\langle\!\sn\hskip-3mm\hookplaq\,\right\rangle\!+\!c_4\!\!\left\langle
\!\!\sn\!\!\twoplaq \,\right\rangle\}\nonumber\\ &&\hskip-8mm=\!\!-\frac{a^4}{2}
(c_0\!+\!8c_1\!\!+\!8c_2\!+\!16c_3\!+\!16c_4)\!\!\sum_{x,\mu,\nu}\!\!\Tr(F_{
\mu\nu}^2(x))\!+\nonumber\\&&\hskip-8mm\frac{a^6}{12}(c_0\!+\!20c_1\!\!-\!4c_2
\!+\!4c_3\!+\!64c_4)\!\!\!\sum_{x,\mu,\nu}\!\!\Tr(\cD_{\mu}F_{\mu\nu}(x))^2\!\!
+\nonumber\\&&\!a^6(\frac{c_2}{3}\!+\!c_3)\!\sum_{x,\mu,\nu,\lambda}
\Tr(\cD_{\mu}F_{\mu\lambda}(x)\cD_{\nu}F_{\nu\lambda}(x))+\nonumber\\
&&\!a^6\frac{c_2}{3}\!\sum_{x,\mu,\nu,\lambda}
\Tr((\cD_\mu F_{\nu\lambda})^2)+\cO(a^8)\ .
\label{eq:weisac}
\eea
Sometimes in the literature $c_2$ and $c_3$ are interchanged~\cite{mgp,wei}.
Here we followed the convention of ref.~\cite{luwe} and we have taken the 
liberty of assigning the coefficient $c_4$ to the $2\times2$ plaquette. The 
$<>$ imply summing $\mu\!\neq\!\nu\ (\neq\!\lambda)$, labelling the edges of
the plaquette, with the point $x$ attached to (say) the lower left 
corner~\cite{mgp}. At 
tree-level only the planar loops are considered, $c_2$ and $c_3$ acquire 
non-zero values only at one-loop order, but as was shown by L\"uscher and 
Weisz~\cite{onsh} field redefinitions allow one to put $c_3\equiv0$. For the 
LW Symanzik action ($c_4=0$) one has $c_0=5/3$ and $c_1=-1/12$ at tree-level, 
but this does not allow for a ``covariant'' gauge condition that will make the 
gauge field propagator diagonal in the space-time indices. The $2\times2$ 
plaquette allows one to ``complete a square'' when choosing $c_4\cdot c_0=
c_1^2$, leading to the gauge fixing functional ($z\equiv c_1/c_0$)
\be\cF_{gf}\equiv\sqrt{c_0}\sum_{\mu}\partial_\mu^\dagger
\left(1\!+\!z(2\!+\!\partial_\mu^\dagger)(2\!+\!\partial_\mu)\right)\!q_\mu(x).
\ee
We decided for this reason to call it the square Symanzik action. Here 
$\partial_\mu$ denotes the lattice difference operator $\partial_\mu\varphi(x)
\equiv \varphi(x+\hat\mu)-\varphi(x)$. As a bonus we note that the condition 
$c_4 c_0=c_1^2$ is invariant under multiplicative link renormalization, as 
they appear in the tadpole improvement scheme, allowing one to easily include 
such factors in a perturbative calculation. At tree-level one now finds 
$c_0=16/9$, $c_1=-1/9$ and $c_4=1/144$. We have verified that this action 
satisfies the positivity bound~\cite{onsh}. It is amusing to see the 
expression for the $a^8$ term in the expansion of the action simplify to
\be
S\!\!=\!\!-\!\!\!\!\sum_{x,\mu,\nu}\!\!\Tr[\frac{a^4\!\!}{2}F_{\mu\nu}^2(x)\!
-\!\frac{a^8\!}{90}(\cD_\mu^2 F_{\mu\nu}(x))^2]\!+\!\!\cO (a^{\!10})
\ee
(behaving as the Symanzik action for cooling~\cite{mgp}).

At tree-level a tadpole parameter $u_0$ modifies $z=c_1/c_0=-1/16$ to 
$z=-1/16u_0^2$, whereas $c_0=1/(1+4z)^2$. One finds in the covariant 
gauge the ghost ($P$) and vector ($P_{\mu\nu}$) propagators to be
\bea
P(k)&=&\!\!\frac{1}{\sqrt{c_0}\sum_\lambda\left(4\sin^2(k_\lambda/2)
+4z\sin^2k_\lambda\right)},\nonumber\\P_{\mu\nu}(k)&=&\!\!
\frac{P(k)\delta_{\mu\nu}}{\sqrt{c_0}\left(1+4z\cos^2(k_\mu/2)\right)}\ .
\eea

\section{EFFECTIVE ACTION}
One can now perform a background field calculation to determine the one-loop
effective action for the zero-momentum gauge fields. For the Wilson action 
and the continuum this was done previously~\cite{vb} and shown to lead to 
rather accurate results. Like for the Wilson case one writes 
$U_\mu(x)=e^{\hat q_\mu(x)}e^{c_\mu(t)/N}$, with $N$ the number of lattice 
sites in the spatial direction (taking the number of sites in the time 
direction infinite) and $\hat q_\mu(x)$ the quantum field, restricted to 
non-zero (spatial) momentum, to be integrated out. This choice
on splitting off the quantum component of the lattice field yields a 
particularly simple background gauge fixing function 
\be
\hat\cF_{gf}\equiv\sqrt{c_0}\sum_{\mu}\!\hat D_\mu^\dagger\!\left(1\!+\!z(2\!
+\!\!\hat D_\mu^\dagger)(2\!+\!\!\hat D_\mu)\right)\!\hat q_\mu(x),
\ee
``covariantizing'' the difference operator to
\be
\hat D_\mu\varphi(x)\equiv e^{c_\mu(t)/N}\varphi(x+\hat\mu)e^{-c_\mu(t)/N}-
\varphi(x).
\ee
\subsection{The SU(2) effective potential}
The one-loop calculation greatly simplifies for an abelian constant background 
field as this allows one to diagonalize the propagator with respect to the
isospin neutral and charged decomposition of the gauge and ghost fields.
The momenta in the background field $\vec c=\half i\vec C\sigma_3$ are shifted,
$\vec k\ra\vec k\!+\!s\vec C/N$, where $s=0$ for
the neutral isospin component and $s=\pm1$ for the two charged components.
It is not hard to find the eigenvalues of the fluctuation operators for the
ghost and gauge fields
\bea
\lambda_{gh}(k)\!\!\!\!&=\!\!\!\!&\sqrt{c_0}\sum_\nu 4\sin^2(k_\nu/2)(1\!+
\!4z\cos^2(k_\nu/2)),\nonumber\\\lambda_\mu(k)\!\!\!\!&=\!\!\!\!&\sqrt{c_0}(
1\!+\!4z\cos^2(k_\mu/2))\lambda_{gh}(k).
\eea
These eigenvalues can be written as products of factors $4\sin^2(k_0/2)+
\omega_\alpha^2$, where the $\omega_\alpha$ can occur in complex conjugate 
pairs at spatial 
momenta close to the edge of the Brillouin zone. As it is well-known that 
the sum over $k_0$ for one such a factor can be performed explicitly,
it is not surprising we can find a closed expression for the effective 
potential, as a sum over the appropriately weighted logarithm of 
the eigenvalues 
\bea
&&\hskip-7mm V^{\rm ab}_1(\vec C)\!=\!N\!\!\!\!\sum_{\vec n\in Z_N^3}\!\!\!
\Bigl\{\sum_i\log\left(\!1\!+\!4z\cos^2\!\left[\frac{2\pi n_i\!+\!C_i}{2N}
\right]\right)\nonumber\\&&\hskip-3mm
+4{\rm asinh}\!\!\left(\!\!2u_0\sqrt{1\!+\!4z\!+\!\!\frac{\omega^2}{2}\!+
\!\omega\sqrt{1\!+\!\frac{\omega^2}{4}}}\right)\!
\Bigr\},
\eea
with $\omega^2\!\!\equiv\!\!4\sum_i\sin^2(k_i/2)\left(1\!+\!4z\cos^2(k_i/2)
\right)$, and $\vec k\!=\!(2\pi\vec n\!+\!\vec C)/N$.
This effective potential, normalized to $V(\vec 0)=0$, is plotted for 
$u_0=1$ in fig.~1 as compared to the result for the Wilson action 
($z\equiv 0$) and for the continuum ($N\ra\infty$).
Although this effective potential is not spectral, since near $\vec C\in2\pi
\zahlen^3$ the adiabatic approximation for integrating out the ``charged''
zero-momentum modes breaks down, one sees that improvement is quite efficient
in removing scaling violations (only scaling violations to fourth order in the 
lattice spacing $a=1/N$ remain). At $N=6$ we can not distinguish the result 
from the continuum at the scale of this figure. One might even fear that 
choosing $u_0\neq1$ makes the agreement worse. 
\begin{figure}[htb]
\vspace{3.6cm}
\includegraphics{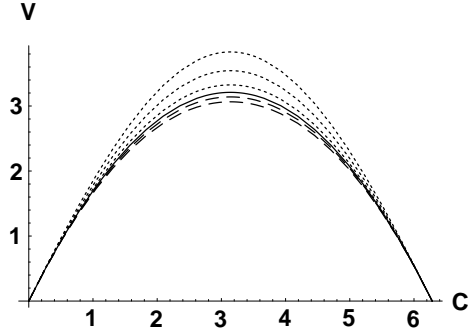}
\caption{The effective potential for a constant Abelian background field 
$c_1=\half iC\sigma_3$. The full line represents the continuum result. The 
lower two dashed curves are for the square Symanzik action with $N=3$ and 4. 
The upper three dotted curves are for the Wilson action with $N=3,4$ and 6.}
\label{fig:veff}
\end{figure}

\subsection{The Lambda ratios}
One can proceed as in the Wilson case with computing the one-loop coefficients
for the effective action~\cite{vb}. In particular this provides for 
$N\ra\infty$ the renormalization of the bare lattice coupling at fixed physical 
volume. As Lorentz invariance is replaced by cubic invariance (both by the 
lattice discretization and the periodic boundary conditions), two independent 
determinations of the Lambda ratios can be extracted from the effective
Lagrangian
\be
\half(\frac{1}{g^2}\!+\!\alpha_1)\!\left(\!\frac{dc^a_i}{dt}\right)^2\!\!\!\!+\!
\quart(\frac{1}{g^2}\!+\!\alpha_2)\!\left(F_{ij}^a\right)^2\!\!\!+\!V_1(c).
\ee
Here $g^{-2}\!\!=\!\!g_0^{-2}\!\!-\!11\log(N)/12\pi^2$ is kept fixed while 
taking the continuum limit. In this limit $\alpha_1$ and $\alpha_2$ differ by 
{\em identical} finite amounts from what was found for the continuum and the 
Wilson action. This finite difference allows one to accurately compute the 
Lambda parameter ratios. These ratios were also computed using the heavy quark 
potential method of ref.~[8], which allows one in addition to obtain the result
for SU(3). For the ratio of the square Symanzik action to the Wilson action we 
find.
\be
\Lambda_{S^2}/\Lambda_W=\pmatrix{4.0919901(1)\quad{\rm for\ SU(2)}\cr
5.2089503(1)\quad{\rm for\ SU(3)}\cr}.
\ee
\subsection{One-loop improvement}
Tests of tadpole corrections to variant tree-level improved actions have been 
performed before~\cite{corn}. However, we remind the reader that there is only 
one pure gauge improved lattice action that was computed to one-loop 
order~\cite{luwe}. It is our aim to bring the square Symanzik action to this 
same level. In principle this provides a way of testing to what extent the 
success of the tadpole improvement depends on the choice of action. 
Independently it is a useful check on the consistency of the Symanzik 
improvement scheme with its inherent redundancy in choosing the lattice 
action to cancel scaling violations. For these perturbative calculations
of the one-loop corrections $c_i^\ol$ one follows the well established 
route of using the twisted finite volume spectroscopy~\cite{luwe}. As a 
normalization condition on the definition of the coupling constant one 
imposes $c_0^\ol+8c_1^\ol+8c_2^\ol+16c_4^\ol=0$. Requiring the physical 
mass of the lowest state to have no quadratic scaling violations to one-loop 
order for the square Symanzik action leads to
\be 
c_1^\ol-c_2^\ol+4c_4^\ol=\pmatrix{-0.00838(1)\quad{\rm for\ SU(2)}\cr
-0.01545(2)\quad{\rm for\ SU(3)}\cr}.
\ee
As an independent check this combination was also extracted (at higher accuracy)
from the heavy quark potential. In addition the on-shell three 
point coupling extracted in the twisted finite volume allows one to 
find~\cite{luwe} a value for $36(c_1^\ol-c_2^\ol+4c_4^\ol)+8c_2^\ol$.
This computation is rather involved and still in progress. Note that $c_4^\ol$
appears in the combinations $c_1^\ol+4c_4^\ol$ and $c_0^\ol-16c_4^\ol$, as is
also dictated by eq.~(2). Therefore as was to be expected $c_4^\ol$ is a free
parameter. It {\em need not}, but can, be fixed by requiring $c_4c_0=c_1^2$ to
one-loop order. Finally we quote the result for the single plaquette 
expectation value: $u_0^4=1\!-\!0.35878\cdot g_0^2(N\!-\!N^{-1})/4$. 

\section{MONTE CARLO RESULTS}
\begin{figure}[htb]
\vspace{7.4cm}
\includegraphics{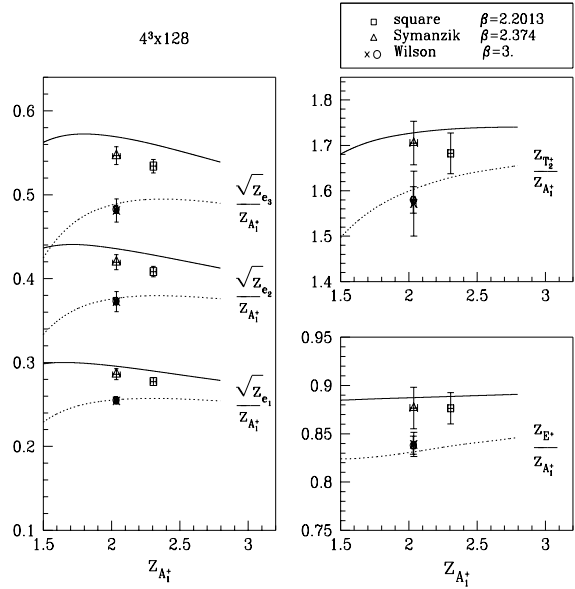}
\caption{SU(2) Monte Carlo data for mass ratios in a small volume on a
lattice of size $4^3\times128$, using the Wilson action (crosses
from existing data of Michael), the LW Symanzik action and our new square
Symanzik action. The lines give the analytic results: full for the continuum
and dotted for the standard lattice action ($N=4$).
\vskip-2mm}
\label{fig:MCdata}
\end{figure}
In the intermediate volume context we can consider at most lattice spacings 
up to 0.25 fermi (asking to be absolved for pushing in this direction), as the 
larger plaquettes that appear in the improved actions require the volume to 
be at least three lattice spacings in each direction. Despite the appearance 
of unitarity violations at the scale of the cutoff, due to the non-local nature
of an improved action, the intermediate volume physical masses remain small 
enough in lattice units to extract them from the decay of correlation functions 
in the time direction in the usual way. In larger volumes and coarser lattices 
the latter problem was dealt with by taking the lattice much finer in the time 
direction~\cite{morn}, using the asymmetric couplings well-known from finite 
temperature studies~\cite{kar}. Here we will consider only Monte Carlo data at 
0.018 fermi ($\beta=4/g_0^2=3$ for the Wilson action, for which we compare our 
data to those by Michael~\cite{mic}). Odd as it may seem, this is 
where the scaling violations for a lattice of 4 lattice spacings in the 
spatial directions are largest within the finite volume spectroscopy~\cite{vb}. 
The data corresponding to the LW Symanzik action is represented by the 
triangles and for our new square Symanzik action by the squares. In both cases 
we used tree-level improvement only. The improvement is significant. For the 
LW Symanzik action the data is within two sigma of the continuum values. The 
results seem to indicate that the square Symanzik action is somewhat less 
effective, although the difference is not significant. A comparison at coarser 
lattices will be more interesting as one should like to see, as advocated, 
tadpole corrections to further improve the results. For this purpose we 
present data elsewhere at a lattice spacing of 0.12 fermi.

\section{HAMILTONIAN}
\subsection{Toy model}
A well-known problem of improved actions is that the transfer matrix is
not hermitian~\cite{ham}. This is easily seen to be related to the 
next-to-nearest neighbor couplings in the time direction. We will illustrate
things here by means of a simple one dimensional problem. For the action
$S(x)=\sum_t(x(t\!+\!1)\!-\!x(t))^2/(2g^2a)\!+\!aV(x(t))$ the partition 
function at finite Euclidean time ($T\!=\!aN$) can be exactly rewritten in 
operator form~\cite{vb}
\be
\cZ=\int\cD xe^{-S(x)}=\Tr(e^{-\half aK}e^{-aV}e^{-\half aK})^N,
\ee
where $K=-\half g^2(\partial/\partial x)^2$. The Hamiltonian read off from 
this equation is only determined up to a unitary transformation. To lowest 
order one finds
\be
H=K+V-\frac{a^2}{24}[V,[K,V]]+\cO(a^4).
\ee
Note that $[V,[K,V]]=g^2 V^\prime(x)^2$.

Next improve the kinetic term $(x(t\!+\!1)\!-\!x(t))^2$ by
$4(x(t\!+\!1)\!-\!x(t))^2/3-(x(t\!+\!2)\!-\!x(t))^2/12$.
One finds that the propagator factorizes as $P(k)\!=\!(P_-(k)\!-\!P_+(k))/Z$, 
where $P^{-1}_\pm\!(k)\!=\!4\sin^2\!(\half k)\!+\omega_\pm^2$, 
$Z\!=\!\sqrt{1\!-\!a^2m^2/3}$, $\omega^2_\pm\!=\!6(1\pm Z)$ and $m^2\!=
\!g^2V''(0)$. This explicitly exhibits the unphysical pole mentioned before 
with masses $m_+^2\sim 12/a$ at the scale of the cutoff. They are not harmful 
for low-energy behavior~\cite{ham}. It would perhaps be misleading to 
associate the spurious poles with ghosts as they do not just occur in loops. 
Vertices do not preserve ghost number. Nevertheless we expect their 
contribution to low-lying states to be suppressed in a way similar to the 
influence of virtual processes due to heavy particles.

Let us introduce the following field redefinition, best expressed 
in the Fourier representation
\be
\bar x(k)\!=\!x(k)\sqrt{1\!+\!\tw\hat k^2}\!-\!\frac{a^2g^2}{24}
\frac{\partial V(\bar x)}{\partial\bar x(-k)},
\ee
where as usual $\hat k=2\sin(k/2)$.
When substituting this non-local transformation in the action we find
\bea
S&\!\!\!=&\!\!\!\!\sum_k\frac{\hat k^2(1\!+\!\tw\hat k^2)
|x(k)|^2}{2ag^2}\!+\!aV(x)\\
&\!\!\!=&\!\!\!\!\sum_k \frac{\hat k^2|\bar x(k)|^2}{2ag^2}\!+\!aV(\bar x)
\!+\!\frac{a^3g^2}{24}V^\prime(\bar x)^2\!\!+\!\cO(a^5),\nonumber
\eea
for which it is assumed that $\hat k=\cO(a)$. We note that after
the field redefinition, ignoring the $\cO(a^5)$ corrections, the action is
local in time and one obtains $H\!=\!K\!+\!V\!+\!\cO(a^4)$ from eq.~(14). 

However, interactions will give rise to a non-trivial Jacobian under this 
change of variables, $J(x)=\det(\partial \bar x(k)/\partial x(k^\prime))$, or
\be
J^2(x)=\det\!\left(1\!+\!\tw\partial^\dagger_0\partial_0\!-\!\tw a^2g^2
\frac{\partial^2V(x)}{\partial x^2}\right),
\ee
where we took the liberty of modifying the $\cO(a^4)$ terms in the operator
whose determinant is to be evaluated. We could likewise define the 
transformation such that the Jacobian is given as above, although this leads
in multi-dimensional cases to non-integrable transformations.

Remarkably one can rewrite this Jacobian up to an $x$ independent factor as
\be
J^2(x)=\det\!\left(1\!-\!a^2 g^2 P_+[\frac{\partial^2V(x)}{\partial x^2}\!
-\!m^2]\right).
\ee
Its contribution to the partition function $\cZ$ can be interpreted as the 
effective action in a background field calculation, with the propagator 
truncated to the unphysical branch, albeit to lowest non-trivial order in the 
lattice spacing. One easily verifies that for $V(x)=\lambda x^4$ this Jacobian 
gives rise to a mass correction {\em linear} in the lattice spacing, which was 
initially discovered by computing the mass gap to first order in $\lambda$ from 
the Feynman rules for the improved action of this simple model. As the model 
is quite similar to the effective action we discussed before, we were quite 
puzzled by this result and it prompted the above derivation.

\subsection{Gauge model}
Indeed, taking the results of sect.~3.1 we can compute easily part of the
effective action for the zero-momentum gauge fields. To obtain the effective 
potential that is valid near $\vec C=\vec 0$, where the tree-level potential 
is quartic in the gauge fields, one restricts the sum to $\vec n\neq\vec 0$ 
and replaces $C_i$ by $r_i\equiv\sqrt{-2\Tr c_i^2}$ in eq.~(9). The result is 
denoted by $\hat V_1(\vec r)$. An accurate description of the full effective 
potential to one-loop order is given by
\be
V_1(c)\!=\!\hat V_1(\vec r)\!+\!\alpha_3r_i^2{F_{jk}^a}^2\!+\!\alpha_4r_i^2
{F_{ij}^a}^2\!+\!\alpha_5{\det}^2\!c
\ee
As can be extracted from the zero-momentum part of eq.~(9), 
$\gamma_1(N)\!=\!\gamma_1(\infty)\!+\!\tw(\sqrt{3}\!-\!1)/N\!+\!\cO(1/N^3)$,
where $\gamma_1$ is the coefficient of $\vec r^{\ 2}$ in the effective 
potential. Generalizing the analysis of the toy model to the situation at 
hand one finds from the Jacobian $\delta_J\gamma_1=-\tw\sqrt{3}/N$. The 
missing piece is provided by the non-triviality of the Haar measure for 
integration over the background link variables
\be
\delta_HV_1(c)=-2N\sum_i\log[2N\sin(r_i/2N)/r_i],
\ee
Indeed one finds $\delta_H\gamma_1=\tw N^{-1}$. Both the Jacobian and
measure contributions compensate for the scaling violations linear in the 
lattice spacing and with it $\hat V_1(\vec r)$ becomes free of scaling 
violations to third order in the lattice spacing. 

Furthermore, rescaling $c$ with $(1\!+\!\tw g^2\gamma_1/N^2)$ removes
to a high degree of accuracy unwanted scaling violations in $\alpha_1$ and
$\alpha_2$. More surprising was to find that the field redefinition 
$\delta c_i=-2g^2\log(N)\cD_\mu^\dagger F_{\mu i}/(24\pi N)^2$ is required to 
remove $\log(N)/N^2$ scaling violations in $\alpha_3$ and $\alpha_5$. As a 
non-trivial check the one-loop coefficient $\alpha_0$, in front of the term 
$\half(\partial_0^\dagger\partial_0c_i^a)^2$, was computed. Its $\log(N)/N^2$ 
term combines {\em after} the above field redefinition with the tree-level 
coefficient of $1/(12g_0^2N^2)$ in precisely the right way to renormalize
the coupling constant.

Remaining $\cO(N^{-2})$ scaling violations will and can be cancelled by the 
one-loop improvement coefficients. Due to the unfortunate mixing with
coefficients that are not easily accessible in lattice perturbation 
theory~\cite{vb}, we cannot at present get at these one-loop improvement 
coefficients along this route. As we have seen, using an effective action
one imposes improvement only up to field redefinitions (or up to unitary 
transformations at the Hamiltonian level). This is more difficult
than computing a few spectral quantities, but has the obvious benefit of 
manifestly improving infinitely many levels at the same time and would be 
a very non-trivial check on improvement indeed. 


\section*{Acknowledgements}

We are grateful to Mark Alford, Tim Klassen, Aida El-Khadra, Sergio Caracciolo,
Mike Teper and in particular Colin Morningstar for discussions. This work was 
supported in part by grants from FOM and from NCF for the use of the Cray C98 
at SARA.


\begin{thebibliography}{9}
\bibitem{sym}K. Symanzik, \Journal{\NPB}{226}{187, 205}{1983}.
\bibitem{luwe}M. L\"{u}scher and P. Weisz, \Journal{\PLB}{158}{250}{1985};
\Journal{\NPB}{266}{309}{1986}.
\bibitem{onsh}M. L\"{u}scher and P. Weisz, \Journal{\CMP}{97}{59}{1985}.
\bibitem{par}G. Parisi, in {\em High Energy Physica-1980}, eds. L. Durand and
L.G. Pondrom (American Institute of Physics, New York, 1981).
\bibitem{lema}G.P. Lepage and P.B. Mackenzie, \Journal{\PRD}{48}{2250}{1993}.
\bibitem{per}V. Periwal, \Journal{\PRD}{53}{2605}{1996}.
\bibitem{mgp}M. Garc\'{\i}a P\'erez, e.a., \Journal{\NPB}{413}{535}{1994}.
\bibitem{wei}P. Weisz, \Journal{\NPB}{212}{1}{1983}; P. Weisz and R. Wohlert,
\Journal{\NPB}{236}{397}{1984}.
\bibitem{vb}P. van Baal, \Journal{\PLB}{224}{397}{1989};
\Journal{\NPB}{351}{183}{1991}.
\bibitem{corn}M. Alford, e.a., \Journal{\PLB}{361}{87}{1995}.
\bibitem{morn} 
C. Morningstar and M. Peardon, \Journal{\NPB}{(Proc.Suppl.) 47}{258}{1996};
hep-lat/\break9606008.
\bibitem{kar}F. Karsch, \Journal{\NPBp}{205[FS5]}{285}{1982}.
\bibitem{mic}C. Michael, \Journal{\NPB}{329}{225}{1990}.
\bibitem{ham}M. L\"{u}scher and P. Weisz, \Journal{\NPB}{240[FS12]}{349}{1984}.
\end{thebibliography}
\end{document}